\documentclass[a4paper]{article}
\usepackage{amsmath,amssymb,amsthm}
\usepackage[utf8]{inputenc}
\usepackage[T1]{fontenc}
\usepackage{epsfig}
\usepackage{booktabs}
\usepackage{hyperref}
\usepackage[backend=biber,eprint=true,sorting=none]{biblatex}
\usepackage{pdflscape}

\newcommand{\bohr}{\mathrm{B}}
\newcommand{\setofreals}{\mathbb{R}}
\newcommand{\setofints}{\mathbb{Z}}
\newcommand{\setintersection}{\cap}
\newcommand{\dualphi}{\tilde{\varphi}}
\newcommand{\dualpsi}{\tilde{\psi}}
\newcommand{\dualeta}{\tilde{\eta}}
\newcommand{\dualzeta}{\tilde{\zeta}}

\newcommand{\projop}[1]{{P^{(#1)}}}
\newcommand{\embeddingop}[1]{{E^{(#1)}}}
\newcommand{\oppartw}[2]{{W^{(#1,#2)}}}
\newcommand{\oppartwind}[3]{{W^{(#1,#2)}_{#3}}}
\newcommand{\opj}[1]{{J^{(#1)}}}
\newcommand{\opjwithind}[3]{{J^{(#1)}_{{#2},{#3}}}}
\newcommand{\oppartfw}[1]{{U^{(#1)}}}

\newcommand{\gtilde}{\tilde{g}}
\newcommand{\abs}[1]{\vert{}#1\vert}
\newcommand{\norm}[1]{\Vert{}#1\Vert}

\newcommand{\szdualappl}[2]{\left\langle #1 , #2 \right\rangle}
\newcommand{\onedimbasisfn}[3]{\psi_{{#1},{#2},{#3}}}
\newcommand{\onedimdualbasisfn}[3]{\dualpsi_{{#1},{#2},{#3}}}

\newcommand{\mdimbasisfn}[3]{\psi_{{#1},{#2},{#3}}}
\newcommand{\mdimdualbasisfn}[3]{\dualpsi_{{#1},{#2},{#3}}}
\newcommand{\mdimgridbasisfn}[1]{{\zeta_{#1}}}
\newcommand{\mdimgriddualbasisfn}[1]{{\dualzeta_{#1}}}
\newcommand{\submin}{\mathrm{min}}
\newcommand{\submax}{\mathrm{max}}
\newcommand{\jmin}{{j_\submin}}
\newcommand{\jmax}{{j_\submax}}
\newcommand{\fnj}{\iota}
\newcommand{\fnjp}{\iota'}

\newcommand{\msci}[2]{{#1} \times 10^{#2}}
\newcommand{\opwithind}[3]{{#1}^{#2}_{#3}}
\newcommand{\veck}{\mathbf{k}}
\newcommand{\vecr}{\mathbf{r}}
\newcommand{\vecrp}{\mathbf{r}'}

\newcommand{\etal}{\textit{et al.}}

\newcommand{\wf}{\phi}                                   % \varphi  olisi parempi (tai \Phi tai \phi) (muutettu)
\newcommand{\nucleusloc}[1]{\mathbf{R}_{#1}}
\newcommand{\entotal}{E_{\textrm{total}}}
\newcommand{\zpm}[1]{Z(#1)}
\newcommand{\zpmthreed}[1]{(Z(#1))^3}
\newcommand{\basisref}[1]{#1}
\newcommand{\spacingsym}{g}
\newcommand{\hartreefock}{Hartree--Fock}

\DeclareMathOperator{\erf}{erf}

\begin{document}

\title{Electronic structure calculations %
  with interpolating tensor product wavelet basis}

\author{Tommi H\"oyn\"al\"anmaa\thanks{corresponding author; email \href{mailto:tommi.hoynalanmaa@tuni.fi}{tommi.hoynalanmaa@tuni.fi}}
  \hspace{2mm}  and Tapio T. Rantala \thanks{email \href{mailto:tapio.rantala@tuni.fi}{tapio.rantala@tuni.fi}}
  \\
  Computational Physics, \\
  P. O. Box 692, FI-33014 Tampere University, Finland}

\maketitle

\begin{abstract}
  We introduce a basis set consisting of three-dimensional
  Deslauriers--Dubuc wavelets and solve numerically the Schrödinger
  equations of H and He atoms and molecules $\mathrm{H}_2$,
  $\mathrm{H}_2^+$, and $\mathrm{LiH}$ with
  \hartreefock\ and DFT methods.  We also compute the \(2\mathit{s}\)
  and \(2\mathit{p}\)
  excited states of hydrogen.  The Coulomb singularity at the nucleus
  is handled by using a pseudopotential.  The eigenvalue problem is solved with
  Arnoldi and Lanczos methods, Poisson equation with GMRES and
  CGNR methods, and matrix elements are computed using the
  biorthogonality relations of the interpolating wavelets.
  Performance is compared with those of CCCBDB and BigDFT. 
\end{abstract}

This article has been accepted for publication in Physical Review E, see
\url{https://journals.aps.org/pre/}.

\section{Introduction}

Standard approaches to assess properties of atoms, molecules and
models of nanostructures
in quantum chemistry are Hartree--Fock (HF) and Density Functional Theory (DFT).
Both of these invoke numerical solutions of the Schr\"odinger
differential equation
of the many-body system of electrons.  Thus, controlled approximations
are inevitable
and practical numerical algorithms are necessary.

Vast majority of the algorithms are based on finding the solutions or
orbitals as series expansion
of basis functions, the basis set.  A finite basis set leads to the
Roothaan--Hall equations, a generalized matrix eigenvalue problem.
Gaussian type basis functions are the most popular, due to their advantageous
analytical features.  Here, we introduce a different type basis set,
wavelet functions, point out
their advantages and drawbacks, and consider a few test cases and
compare their performance with conventional approaches.

%% In principle, properties of a quantum physical system are computed by
%% solving its Schr\"odinger equation. However, the Schr\"odinger
%% equation usually has to be approximated for real world computations
%% with several particles. The approximate solution is often computed
%% with \hartreefock\ (HF) or Density Functional Theory (DFT)
%% method. Using the HF method and a finite basis set leads to the
%% Roothaan--Hall equations that are a generalized matrix eigenvalue
%% problem.  Wavelets can be used as a basis set for the HF and DFT
%% methods.

Wavelets and related scaling functions are functions generated by
translations
and dilatations of the so called mother wavelet and mother scaling
function.
Interpolating wavelets use a mother scaling function satisfying the cardinal
interpolating property \(\varphi(k) = \delta_{k,0}\) where \(k\) is an
integer. Orthonormal wavelets form an orthonormal basis in function
space \(L^2(\setofreals)\). Both of these wavelet types can be
generalized to multivariate functions.

Studies of wavelets have been active and new ones with various
properties have been found during the last tens of years
\cite{daubechies1992,cl1996}.
One-dimensional interpolating
wavelets in function space \(C_u(\setofreals)\) consisting of bounded
and uniformly continuous functions on \(\setofreals\) are defined in
Ref. \cite{cl1996}. 
One-dimensional interpolating
wavelets in function space \(C_0(\setofreals)\) consisting of
functions on \(\setofreals\) vanishing at infinity are defined in
Ref. \cite{donoho1992}.
Deslauriers--Dubuc wavelets have also been discussed in
Refs. \cite{dd1989} and \cite{dubuc1986}.
Compactly supported interpolating wavelets have been generalized to
multiple dimensions in Refs. \cite{hoynalanmaa2015} and \cite{goedecker1998}.
Fukuda, Kinoshita, and Suzuki \cite{fks2016} have studied
unconditional convergence of wavelet expansions. They have shown that
uniformly convergent wavelet expansions even for continuous
functions do not always converge unconditionally in
\(L^\infty(\mathbb{R})\).
Pathak \cite{pathak2011} has investigated translation and convolution
associated with discrete wavelet transform.

Arias \cite{arias1999} and Engeness and Arias \cite{ea2002} have developed
formalism for electronic structure calculations with interpolating
wavelets so that matrix elements of the operators are computed as
usual and overlap matrices are used in the matrix form of the
Schr\"odinger equation.  Lippert \etal\ \cite{lae1998} introduce a
Lagrangian based formalism for the multiresolution analysis (MRA) of
electronic structure.  Arias \cite{arias1999} uses carbon atom and
\(\mathrm{N}_2\) molecule as examples. Engeness and Arias
\cite{ea2002} use both calcium and aluminum atoms and molecules \(\mathrm{O}_2\) and
\(\mathrm{H}_2\mathrm{O}\)  as examples.

Fischer and Defranceschi
use Daubechies wavelets \cite{daubechies1992} for computation of
hydrogenlike atoms \cite{fd1998}.  They have developed an iterative method
based on nonstandard operator form of the Schr\"odinger
operator. Their work shows that this method is well suited for
computations of hydrogenlike atoms.
Fischer and Defranceschi have
presented \hartreefock\ equations in an orthonormal wavelet basis
\cite{fd1994a}.  They have also analyzed the \hartreefock\ method with
continuous wavelet transform \cite{fd1994b} and demonstrated it using
the hydrogenic Schrödinger equation with an iterative solution scheme.  

Wei and Chou \cite{wc96}
have used orthonormal wavelets in self-consistent electronic structure
calculations within the local-density approximation and demonstrated
their method with
\(\mathrm{H}_2\) and \(\mathrm{O}_2\) molecules.  Tymczak and Wang
\cite{tw97} have used orthonormal Daubechies wavelets for quantum molecular
dynamic simulations and developed a wavelet selection scheme for
computations. They used hydrogen atom and \(\mathrm{H}_2\)
molecule as examples. Their method shows systematic convergence with
increasing grid size.  Yamaguchi and Mukoyama \cite{ym1997} have carried out
electronic structure calculations with the \hartreefock\ method and Meyer wavelets.

Our earlier contribution  \cite{hrr2004} is use of one-dimensional
interpolating wavelets to solve the \hartreefock\ equations in the 
Central Field Approximation for orbitals of several test case atoms.
We were able to derive analytic formulas for all the relevant 
matrix elements of Hamiltonian and Fock operators. We also developed
the Exact Pseudopotential Method \cite{atoms11010009} for
one-dimensional calculations of atoms and applied it to the hydrogen
and helium ground states and for some excited states.

%There are not so many comparative studies of use of wavelet basis and the conventional Gaussian basis.   Onko näin?
Iyengar and Frisch \cite{if2004} have studied relationship between
Gaussian basis sets and wavelets.  They use a time-dependent basis
function set: Gaussian functions centered at nuclei of the
system. When the nuclei move the basis functions move, too.
Gaussian functions are an example of \textit{multiwavelets}
\cite{tnr2002,strang1994orthogonal,if2004}, for which the $J$th level
scaling space is decomposed into $N$ scaling spaces via
\begin{equation}
  \ldots \mathbf{V}^{(N)}_{J-1} \subset \mathbf{V}^{(1)}_J \subset
  \mathbf{V}^{(2)}_J \subset \cdots \subset \mathbf{V}^{(N)}_J
  \subset \mathbf{V}^{(1)}_{J+1} \ldots .
\end{equation}
Comparative studies of wavelets
and Gaussian functions are presented e.g. in
\cite{gossler2021gaussian} and \cite{singh2016comparative}.
% tässä paperissa on enemmänkin mainitsemista: "time-dependent basis functions"???
Harrison
\etal\ \cite{harrison2003multiresolution} use multiwavelets for
quantum chemistry computations. Yanai
\etal\ \cite{yanai2004multiresolution} present a numerical algorithm
to evaluate \hartreefock\ exchange in the SCF method.
Yanai \etal\ \cite{yanai2005multiresolution,yanai2015multiresolution}
develop a method to do time dependent \hartreefock\ and Density
Functional Theory computation with multiwavelets.
Jensen \etal\ \cite{jensen2017elephant} perform multiwavelet
computations of total energies with GGA-PBE and hybrid-PBE0 density
functionals for 211 molecules. Jensen \etal\ \cite{jensen2016magnetic}
compute some magnetic properties with multiwavelets.

Han, Cho, and Ihm \cite{hci1999} have developed an
all-electron density-functional program using the Mexican hat wavelets.
They analyze \(\mathrm{H}_2\), \(\mathrm{CO}\), and
\(\mathrm{H}_2\mathrm{O}\) molecules and \(1\mathit{s}\) core-ionized C*O and
CO* molecules.  Their method shows very good performance over the
plane-wave based methods.
Genovese \etal\ \cite{g2008}
and Mohr \etal\ \cite{m2014} have composed a software package
BigDFT that implements the DFT method for quantum physical systems
using three-dimensional Daubechies wavelets as a basis function set.

In three dimensional space the number of basis functions may grow
relatively large to give sufficient accuracy.  Therefore, we do not construct
the Hamiltonian matrix explicitly.  %Because the number of basis functions is so large 3-dimensional quantum physical problems cannot usually be solved by explicitly constructing the matrix of the Hamiltonian operator. 
Instead, we use  %proper 
iterative algorithms in solving the Roothaan--Hall
equation, though correspondingly, the computation gets slower % Execution time of the solution tends to be slow when there are plenty of basis functions.
% We use 3-dimensional tensor product Deslauriers--Dubuc wavelets to solve the wave equations
in self-consistent iteration of  \hartreefock\ or DFT orbitals 
for many electron systems.

%The new issue in this article is that we compute electronic structure
In this paper, we demonstrate evaluation of electronic structure
with three-dimensional interpolating tensor product wavelets and use
of dual multiresolution analysis in computation of the matrix elements
of the various operators.
We make calculations
for the hydrogen and helium atoms, hydrogen molecule ion, hydrogen
molecule, and lithium hydride molecule.
Self-consistent iteration and \hartreefock\ and DFT methods are used
for many electron systems.
Pseudopotentials are used for two different purposes, to handle the
Coulomb singularity at nuclei, and also, as frozen-core.
%approximate the core electrons so that they need not be calculated in the computation.
We use the data from BigDFT as the reference for 
our calculations with the interpolating 
tensor product Deslauriers--Dubuc wavelet basis function set.

An advantage of the interpolating wavelets compared to the orthonormal
wavelets is that computing a wavelet expansion of the function does
not require numerical evaluation of integrals since the dual wavelets
are weighted sums of delta functions. In case the interpolating
wavelets are compactly supported these sums are finite, too. We have
to choose the wavelet family so that its Hölder regularity is at least
2 in order to enable the evaluation of the Laplacian operator. BigDFT
uses Daubechies orthonormal wavelets for the computation of atomic
orbitals and scalar products and interpolating wavelets for charge
density, function products, and the Poisson solver. We use a
rectangular computation grid whereas BigDFT uses a spherical one.

From now on, we use atomic units throughout this article 
(\(e = m_e = \hbar = 4 \pi \varepsilon_{0} = 1\)).
Thus, units for energy and length, ``Hartree'' and ``Bohr'' are used and abbreviated as
\(\mathrm{Ha}\) and \(\bohr\).
Notations for wavelet basis functions and filters is similar to that in
Ref. \cite{hoynalanmaa2015}
and computation of matrix elements is similar to that in Ref. \cite{goedecker1998}.

\section{Solving the Schrödinger Equations}
\label{sec:schrodinger}

\subsection{General}
\label{sec:schrodinger-general}

Consider a system with \(n\) electrons and \(m\) nuclei,
with atomic numbers \(Z_i\) and locations \(\nucleusloc{i}\). Within the Born--Oppenheimer approximation,
dynamics of electrons and nuclei are independent, and then, the wave function separates to two factors, correspondingly.
% We do not consider the vibrational and rotational states.
Here, we consider the electronic part, the orbitals, only, and keep the nuclear conformation \{\(\nucleusloc{i}\)\}
fixed.
%In this article we are interested in the electronic orbitals only.

With the fixed nuclear conformation, the Coulomb potential for dynamics of electrons is
\begin{equation}
  V_{\textrm{N}}(\vecr) = - \sum_{i=1}^m \frac{Z_i}{\abs{\vecr-\nucleusloc{i}}}
\end{equation}
and the internuclear repulsion energy
\begin{equation}
  E_{\textrm{R}} = \sum_{i=1}^{m-1} \sum_{j > i}^m
  \frac{Z_iZ_j}{\abs{\nucleusloc{i}-\nucleusloc{j}}} .
\end{equation}
For an atom we have \(E_{\textrm{R}} = 0\).

% Mielestäni esim. \varphi olisiparempi kuin \gamma (muutettu)
Let us denote the orbitals by \(\wf\), as symbols
\(\varphi\) and \(\psi\) are used for scaling functions and wavelets.

If the spacing between grid points is a (usually negative) power of
two we can handle this by choosing \(j_\submin\) in equations
\eqref{eq:aac} and \eqref{eq:aad} properly.
Otherwise we have to make a change of variables
\(\mathbf{r} = a \mathbf{r}'\) in the Schrödinger equation. Here one
unit in the computation grid corresponds to \(a\) Bohrs.

% Tämä osa ei ole sujuvasti kirjoitettu (muutettu)
%% If the grid spacing (see section \ref{sec:basis-set}) is not a
%% (usually negative) power of two we have to make a change of variables
%% \(\mathbf{r} = a \mathbf{r}'\) in the Schrödinger equation. Here one
%% unit in the computation grid corresponds to \(a\) Bohrs.

\subsection{Single-electron System}

The Schrödinger equation of a single electron system is
\begin{equation}
    \left( -\frac{1}{2} \nabla^2 + V_{\textrm{N}} \right) \wf = \varepsilon \wf ,
\end{equation}
where $\varepsilon$ is the orbital energy and the total energy
including nuclear repulsion is
\begin{equation}
  \entotal = \varepsilon + E_{\textrm{R}} .
\end{equation}
The wave equation of a single-electron system is solved by the
Implicitly Restarted Arnoldi Method \cite{arnoldi1951,ls1996}. The
Arnoldi method is able to find also other than the lowest eigenvalue.

\subsection{\hartreefock\ Method}

% Tuo kerroin 1/2 Hartree-potentiaalissa on tietysti oikein, mutta
% näyttää huonolta, koska se on totta vain 2-elektronisysteemille,
% jossa molemmat elektronit sattuvat olemaan samalla orbitaalilla.
% (muutettu)

The HF equation for an \(n\) electron system is
\begin{equation}
  \left( -\frac{1}{2} \nabla^2 + V_{\textrm{N}}
  + V_{\textrm{H}} + V_{\textrm{x}}^i \right ) \wf_i
  = \varepsilon_i \wf_i
\end{equation}
where the Hartree potential is given by
\begin{equation}
  V_{\textrm{H}}(\mathbf{r}) = \int_{\setofreals^3}
  \rho(\mathbf{r'})
  \frac{1}{\abs{\mathbf{r}-\mathbf{r'}}} d^3\mathbf{r'}
\end{equation}
and the charge density by
\begin{equation}
  \rho(\vecr) = \sum_{i=1}^n \abs{\wf_i(\vecr)}^2 .
\end{equation}

The exchange potentials \(V_{\textrm{x}}^i\), \(i = 1, \ldots, n\),
are defined in \cite[section 2.2]{saad2010numerical} but not
needed in our study.
For a two electron system with both electrons occupying the same orbital
the singlet state HF equation can be written as
\begin{equation}
  \left( -\frac{1}{2} \nabla^2 + V_{\textrm{N}} + \frac{1}{2}
  V_{\textrm{H}} \right ) \wf_1
  = \varepsilon_1 \wf_1 .
\end{equation}
The Hartree potential is computed by solving the
Poisson equation
\begin{equation}
  \label{eq:poisson}
  \nabla^2 V_{\textrm{H}} = -4 \pi \rho
\end{equation}
numerically.

% Tämä lienee näin vain, jos kanta on aallokkeita? (kyllä)
In case, where there are more than one resolution level in an interpolating
wavelet basis the matrix \(L\) of the Laplacian operator is not
generally Hermitian so we cannot solve equation \eqref{eq:poisson} directly
with the conjugate gradient method.  The nonhermiticity arises because
our matrix elements are not computed as ordinary inner products
between functions.
In this case we use one of the
following two methods:
\begin{itemize}
  \item conjugate gradient on the normal equations (CGNR): solve
    \begin{equation}
      L^T L V_{\textrm{H}} = -4 \pi L^T \rho
    \end{equation}
    with the conjugate gradient method.
  \item generalized minimal residual method (GMRES) \cite{ss1986}.
\end{itemize}
When the basis set consists of a single resolution level we may use
ordinary conjugate gradient method to solve \eqref{eq:poisson}.

The total energy of a two electron system is
\begin{equation}
  \entotal = 2 \varepsilon_1 - \frac{1}{4} \int_{\setofreals^3} \rho(\mathbf{r})
  V_{\textrm{H}}(\mathbf{r}) d^3\mathbf{r}
  + E_{\textrm{R}} .
\end{equation}
In this paper, we consider restricted \hartreefock\ approach, only.

\subsection{Density Functional Theory and Local Density Approximation}

Suppose that we have a system with \(M\) electronic orbitals whose
total wavefunction is \(\Psi\).
The Kohn-Sham equation \cite{kohn1999,saad2010numerical,sx2002} for
the electronic structure is
\begin{equation}
  \left( -\frac{1}{2} \nabla^2 + V_{\textrm{N}} + V_{\textrm{H}} +
  V_{\textrm{xc}}[\rho] \right) \wf_i
  = E_i \wf_i
\end{equation}
where the charge density is
\begin{equation}
  \rho(\mathbf{r}_1) =
  \sum_{s_1= \pm 1/2} M \int  \vert \Psi\left(\mathbf{x}_1, \mathbf{x}_2,
  \ldots, \mathbf{x}_M\right) \vert^2 d\mathbf{x}_2 \ldots d\mathbf{x}_M
\end{equation}
and the Hartree potential
\begin{equation}
  V_{\textrm{H}}(\mathbf{r}) = \int_{\setofreals^3}
  \rho(\mathbf{r'})
  \frac{1}{\abs{\mathbf{r}-\mathbf{r'}}} d^3\mathbf{r'} .
\end{equation}
and \(V_{\textrm{xc}}[\rho]\) is the \textit{exchange-correlation potential}.
We have
\begin{equation}
  V_\textrm{xc}[\rho] = V_\textrm{x}[\rho] + V_\textrm{c}[\rho]
\end{equation}
where \(V_\textrm{x}[\rho]\) is the
\textit{exchange potential} and \(V_\textrm{c}[\rho]\) is the
\textit{correlation potential}. In this article we set
\(V_\textrm{c}[\rho] = 0\). The exchange-correlation energy is defined
by
\begin{equation}
  E_\textrm{xc}[\rho] = E_\textrm{x}[\rho] + E_\textrm{c}[\rho]
\end{equation}
where \(E_\textrm{x}[\rho]\) is the
\textit{exchange energy} and \(E_\textrm{c}[\rho]\) is the
\textit{correlation energy}. In this article we ignore the correlation energy.

% Ahaa, . . ..  No, onhan se loogista, kun ei HF:n päällekään tule korrelaatiota.  Tämän ehkä voisi perustella sillä tavalla.  :)
% Voisi olla kirjoittaa, että "ei oteta huomioon", kuin sanoa, että se on nolla.
% (muutettu)
Within the Local Density Approximation (LDA) we define
\begin{equation}
  E_{\textrm{x}}[\rho] = \int \rho(\vecr) \varepsilon_{\mathrm{x}}[\rho](\vecr) d^3 \vecr
\end{equation}
where \(\varepsilon_{\mathrm{x}}[\rho](\vecr)\) is the exchange energy per
particle of a uniform electron gas at a density of \(\rho\).
It follows from the Kohn-Sham theorem
\cite[section 3.1]{saad2010numerical} that the exchange potential is
\begin{equation}
  V_{\textrm{x}}[\rho] = \frac{\delta E_{\textrm{x}}[\rho]}{\delta \rho} .
\end{equation}
We have
\begin{equation}
  E_{\textrm{x}}[\rho] = -\frac{3}{4} \left( \frac{3}{\pi}
  \right)^{1/3} \int \left( \rho(\vecr) \right)^{4/3} d^3
  \vecr
\end{equation}
and
\begin{equation}
  V_{\textrm{x}}[\rho](\vecr) = - \left(\frac{3}{\pi}
  \rho(\vecr)\right)^{1/3} .
\end{equation}              
% tähän LDA:han ja tuohon vaihtoenergiaan ja -potentilaaliin pitäisi
% kyllä saada viite tai pari (lisätty)

The total energy of the system is
\begin{equation}
  E_{\textrm{KS}} = \sum_{i=1}^n E_i
  - \frac{1}{2} \int \rho(\vecr) V_{\textrm{H}}(\vecr) d^3 \vecr
  + E_{\textrm{xc}}[\rho]
  - \int \rho(\vecr) V_{\textrm{xc}}[\rho](\vecr) d^3 \vecr .
\end{equation}
The Kohn-Sham equations are solved by a similar self-consistent
iteration as the HF equations.

\section{Three-dimensional Wavelet Basis Set}

\subsection{The Basis Set}
\label{sec:basis-set}

Let \( j_\submin \) and \( j_\submax \)
be the minimum and maximum resolution levels of the point grid.
Let
\begin{equation}
  \label{eq:eqaaa}
  Z_j = \left\{ \frac{k}{2^j} \bigg\vert k \in \setofints \right\}
\end{equation}
and
\begin{equation}
  \label{eq:eqaab}
  V_j = Z_j^3
\end{equation}
where \( j \in \setofints \). 
Define sets \(Q_j\) by
\begin{eqnarray}
  \label{eq:aac}
  Q_{j_\submin} & = & V_{j_\submin} \\
  \label{eq:aad}
  Q_j & = & V_j \setminus V_{j-1} \;\; \textrm{for} \; j > j_\submin
\end{eqnarray}
The point grid \(G\) shall be some finite subset of \(V_{j_\submax}\).
We define
\begin{equation}
  \label{eq:gj}
  G_j := G \setintersection Q_j
\end{equation}
for \(j \geq \jmin\).
The functions \(\varphi_{j,k}\) and \(\psi_{j,k}\) are
scaling functions and wavelets belonging to an interpolating
wavelet family. Functions \(\dualphi_{j,k}\) and
\(\dualpsi_{j,k}\) are dual basis functions of interpolating wavelets.

Define
\begin{eqnarray}
  \onedimbasisfn{s}{j}{k} & := &
  \left\{
  \begin{array}{ll}
    \varphi_{j,k} ; & \textrm{ if } s = 0 \\
    \psi_{j,k} ; & \textrm{ if } s = 1 \\
  \end{array}
  \right. \\
  \onedimdualbasisfn{s}{j}{k} & := &
  \left\{
  \begin{array}{ll}
    \dualphi_{j,k} ; & \textrm{ if } s = 0 \\
    \dualpsi_{j,k} ; & \textrm{ if } s = 1 \\
  \end{array}
  \right.
\end{eqnarray}
and
\begin{eqnarray}
  \eta_{j,k} & := &
  \left\{
  \begin{array}{ll}
    \varphi_{\jmin,k} ; & \textrm{ if } j = \jmin \\
    \varphi_{j-1,k/2} ; & \textrm{ if } j > \jmin \textrm{ and } k \textrm{ even} \\
    \psi_{j-1,(k-1)/2} ; & \textrm{ if } j > \jmin \textrm{ and } k \textrm{ odd}
  \end{array}
  \right. \\
  \dualeta_{j,k} & := &
  \left\{
  \begin{array}{ll}
    \dualphi_{\jmin,k} ; & \textrm{ if } j = \jmin \\
    \dualphi_{j-1,k/2} ; & \textrm{ if } j > \jmin \textrm{ and } k \textrm{ even} \\
    \dualpsi_{j-1,(k-1)/2} ; & \textrm{ if } j > \jmin \textrm{ and } k \textrm{ odd}
  \end{array}
  \right.
\end{eqnarray}
When \(\alpha \in Q_j\) and \(j \geq \jmin\) define
\begin{equation}
  \label{eq:mdimgridbasisfn}
  \mdimgridbasisfn{\alpha} :=
  \eta_{j,\veck[1]} \otimes
  \eta_{j,\veck[2]} \otimes
  \eta_{j,\veck[3]}
\end{equation}
and
\begin{equation}
  \mdimgriddualbasisfn{\alpha} :=
  \dualeta_{j,\veck[1]} \otimes
  \dualeta_{j,\veck[2]} \otimes
  \dualeta_{j,\veck[3]}
\end{equation}
where \(\veck = 2^j \alpha\).
We also define
\begin{equation}
  \varphi_{j,\veck} :=
  \varphi_{j,\veck[1]} \otimes
  \varphi_{j,\veck[2]} \otimes
  \varphi_{j,\veck[3]}
\end{equation}
where \(j \in \setofints\) and \(\veck \in \setofints^3\).

\subsection{Backward and Forward Wavelet Transforms}
\label{sec:transforms}

Let
\begin{equation}
  \label{eq:aai}
  f = \sum_{\alpha \in G} c_\alpha \zeta_\alpha
\end{equation}
where \( c_\alpha \in \setofreals \) for all \( \alpha \in G \).
Let \( c = ( c_\alpha )_{\alpha \in G} \).
Define \( v = ( v_\alpha )_{\alpha \in G} \) by setting
\begin{equation}
  \label{eq:aaj}
  v_\alpha = f( \alpha ).
\end{equation}
We define forward wavelet transform \(U\)
and backward wavelet transform \(W\) by setting \( U(v) = c \) and \(
W(c) = v\). Mappings \(U\) and \(W\) are linear. We compute the
forward wavelet transform \(U\) using an algorithm somewhat similar to
\cite{vp96}.
Define matrix \(\projop{j}\) by
\begin{equation}
  \label{eq:aaka}
  \opwithind{P}{(j)}{\alpha,\beta} = \delta_{\alpha,\beta}
\end{equation}
where \(\alpha \in G_j\) and \(\beta \in G\)
and matrix \(\embeddingop{j}\) by
\begin{equation}
  \label{eq:aakb}
  \opwithind{E}{(j)}{\alpha,\beta} = \delta_{\alpha,\beta}
\end{equation}
where \(\alpha \in G\) and \(\beta \in G_j\).
Define
\begin{equation}
  \oppartwind{j}{j'}{\alpha,\beta} = \mdimgridbasisfn{\beta}(\alpha)
\end{equation}
where \(\alpha \in G_j\) and \(\beta \in G_{j'}\). See equation
\eqref{eq:gj} for definition of \(G_j\) and equation
\eqref{eq:mdimgridbasisfn} for definition of \(\mdimgridbasisfn{\beta}\).
We have
\begin{equation}
  \label{eq:aala}
  W = \sum_{j=\jmin}^{\jmax} \sum_{j'=\jmin}^{j} 
  \embeddingop{j} \oppartw{j}{j'} \projop{j'} .
\end{equation}
For forward wavelet transform we have
\begin{eqnarray}
  \label{eq:aama}
  U & = & \sum_{j=\jmin}^{\jmax} \embeddingop{j} \oppartfw{j} \\
  \label{eq:aamb}
  \oppartfw{j} & = & \opj{j} \left( \projop{j} - 
    \sum_{j'=\jmin}^{j-1} \oppartw{j}{j'} \oppartfw{j'} \right), \\
  \nonumber
  & & \textrm{for } j > \jmin \\
  \label{eq:aamc}
  \oppartfw{\jmin} & = & \projop{\jmin} \\
  \label{eq:aamd}
  \opj{j} & = & \left( \oppartw{j}{j} \right)^{-1} 
\end{eqnarray}

When
\begin{equation}
  f = \sum_{\beta \in G_j} c_\beta \zeta_\beta
\end{equation}
we have
\begin{equation}
  c_\alpha = \szdualappl{\dualzeta_\alpha}{f}
  = \szdualappl{\dualzeta_\alpha}{\sum_{\beta \in G_j} f(\beta) \varphi_{j,2^j\beta}}
  = \sum_{\beta \in G_j}
  \szdualappl{\dualzeta_\alpha}{\varphi_{j,2^j\beta}} f(\beta) .
\end{equation}
Consequently 
\begin{equation}
  \opjwithind{j}{\alpha}{\beta} = \szdualappl{\dualzeta_\alpha}{\varphi_{j,2^j\beta}}
\end{equation}
and we do not have to invert matrix \(\oppartw{j}{j}\).

An operator representing pointwise multiplication of a function \(f\) by
another in the given computation grid is
\begin{equation}
  M = UDW
\end{equation}
where \(D\) is a diagonal matrix with values of the function \(f\) at
the grid points in the diagonal. The local pseudopotentials are
computed this way.

\subsection{Matrix Elements of the Laplacian Operator}
\label{sec:laplacian}

Laplacian operator
\begin{equation}
  \label{eq:laplacian}
  \nabla^2 = \frac{\partial^2}{\partial x^2}
  + \frac{\partial^2}{\partial y^2}
  + \frac{\partial^2}{\partial z^2}
\end{equation}
is approximated by linear operator 
\(L = L^{(x)} + L^{(y)} + L^{(z)}\) where
\begin{eqnarray}
  \label{eq:aapa}
  L^{(x)}_{\alpha,\alpha'} & = & \int_{\setofreals^3}
  \dualzeta_{\alpha}(\mathbf{x}) \frac{\partial^2}{\partial x^2} 
  \zeta_{\alpha'}(\mathbf{x}) d\tau \\
  \label{eq:aapb}
  L^{(y)}_{\alpha,\alpha'} & = & \int_{\setofreals^3}
  \dualzeta_{\alpha}(\mathbf{x}) \frac{\partial^2}{\partial y^2}
  \zeta_{\alpha'}(\mathbf{x}) d\tau \\
  \label{eq:aapc}
  L^{(z)}_{\alpha,\alpha'} & = & \int_{\setofreals^3}
  \dualzeta_{\alpha}(\mathbf{x}) \frac{\partial^2}{\partial z^2}
  \zeta_{\alpha'}(\mathbf{x}) d\tau
\end{eqnarray}
for \(\alpha, \alpha' \in G\).
Let \(\alpha = 2^{-j} (k_x,k_y,k_z) \in G \setintersection Q_j \) and
\(\alpha' = 2^{-j'} (k_x',k_y',k_z') \in G \setintersection Q_{j'} \). We define
\begin{eqnarray}
  \label{eq:aaqaa}
  \fnj & = &
  \left\{
    \begin{array}{ll}
      j - 1, & \textrm{ if } j > \jmin \\
      j, & \textrm{ if } j = \jmin
    \end{array}
  \right. \\
  \label{eq:aaqa}
  l_x  & = &
  \left\{
    \begin{array}{ll}
      \frac{k_x}{2}, & \textrm{if } k_x \textrm{ even and } j > \jmin
      \\
      \frac{k_x-1}{2}, & \textrm{if } k_x \textrm{ odd and } j > \jmin
      \\
      k_x, & \textrm{if } j = \jmin
    \end{array}
  \right. \\
  \label{eq:aaqb}
  t_x  & = &
  \left\{
    \begin{array}{ll}
      0, & \textrm{if } j = \jmin \textrm{ or } j > \jmin
      \textrm{ and } k_x \textrm{ even} 
      \\
      1, & \textrm{if } j > \jmin
      \textrm{ and } k_x \textrm{ odd}
    \end{array}
  \right. \\
  \label{eq:aaqc}
  l_y  & = &
  \left\{
    \begin{array}{ll}
      \frac{k_y}{2}, & \textrm{if } k_y \textrm{ even and } j > \jmin
      \\
      \frac{k_y-1}{2}, & \textrm{if } k_y \textrm{ odd and } j > \jmin
      \\
      k_y, & \textrm{if } j = \jmin
    \end{array}
  \right. \\
  \label{eq:aaqd}
  t_y  & = &
  \left\{
    \begin{array}{ll}
      0, & \textrm{if } j = \jmin \textrm{ or } j > \jmin
      \textrm{ and } k_y \textrm{ even} 
      \\
      1, & \textrm{if } j > \jmin
      \textrm{ and } k_y \textrm{ odd}
    \end{array}
  \right. \\
  \label{eq:aaqe}
  l_z  & = &
  \left\{
    \begin{array}{ll}
      \frac{k_z}{2}, & \textrm{if } k_z \textrm{ even and } j > \jmin
      \\
      \frac{k_z-1}{2}, & \textrm{if } k_z \textrm{ odd and } j > \jmin
      \\
      k_z, & \textrm{if } j = \jmin
    \end{array}
  \right. \\
  \label{eq:aaqf}
  t_z  & = &
  \left\{
    \begin{array}{ll}
      0, & \textrm{if } j = \jmin \textrm{ or } j > \jmin
      \textrm{ and } k_z \textrm{ even} 
      \\
      1, & \textrm{if } j > \jmin
      \textrm{ and } k_z \textrm{ odd}
    \end{array}
  \right.
\end{eqnarray}
and similar definitions for 
\(\fnjp, l_x', t_x', l_y', t_y', l_z', \textrm{ and } t_z' \).
Elements of matrices are computed by
\begin{eqnarray}
  \label{eq:aara}
  L^{(x)}_{\alpha,\alpha'} & = &
  \left\{
    \begin{array}{ll}
      2^{2\fnj} a( t_x, t_x', \fnjp-\fnj, l_x'-2^{\fnjp-\fnj}l_x ) & \\
      \cdot s( t_y, t_y', \fnjp-\fnj, l_y'-2^{\fnjp-\fnj}l_y ) & \\
      \cdot s( t_z, t_z', \fnjp-\fnj, l_z'-2^{\fnjp-\fnj}l_z ), &
      \textrm{ if } \fnj \leq \fnjp \\
      2^{2\fnjp} a( t_x, t_x', \fnjp-\fnj, l_x-2^{\fnj-\fnjp}l_x' ) & \\
      \cdot s( t_y, t_y', \fnjp-\fnj, l_y-2^{\fnj-\fnjp}l_y' ) & \\
      \cdot s( t_z, t_z', \fnjp-\fnj, l_z-2^{\fnj-\fnjp}l_z' ), &
      \textrm{ if } \fnj > \fnjp
    \end{array}
  \right. \\
  \label{eq:aarb}
  L^{(y)}_{\alpha,\alpha'} & = &
  \left\{
    \begin{array}{ll}
      2^{2\fnj} s( t_x, t_x', \fnjp-\fnj, l_x'-2^{\fnjp-\fnj}l_x ) & \\
      \cdot a( t_y, t_y', \fnjp-\fnj, l_y'-2^{\fnjp-\fnj}l_y ) & \\
      \cdot s( t_z, t_z', \fnjp-\fnj, l_z'-2^{\fnjp-\fnj}l_z ), &
      \textrm{ if } \fnj \leq \fnjp \\
      2^{2\fnjp} s( t_x, t_x', \fnjp-\fnj, l_x-2^{\fnj-\fnjp}l_x' ) & \\
      \cdot a( t_y, t_y', \fnjp-\fnj, l_y-2^{\fnj-\fnjp}l_y' ) & \\
      \cdot s( t_z, t_z', \fnjp-\fnj, l_z-2^{\fnj-\fnjp}l_z' ), &
      \textrm{ if } \fnj > \fnjp
    \end{array}
  \right. \\
  \label{eq:aarc}
  L^{(z)}_{\alpha,\alpha'} & = &
  \left\{
    \begin{array}{ll}
      2^{2\fnj} s( t_x, t_x', \fnjp-\fnj, l_x'-2^{\fnjp-\fnj}l_x ) & \\
      \cdot s( t_y, t_y', \fnjp-\fnj, l_y'-2^{\fnjp-\fnj}l_y ) & \\
      \cdot a( t_z, t_z', \fnjp-\fnj, l_z'-2^{\fnjp-\fnj}l_z ), &
      \textrm{ if } \fnj \leq \fnjp \\
      2^{2\fnjp} s( t_x, t_x', \fnjp-\fnj, l_x-2^{\fnj-\fnjp}l_x' ) & \\
      \cdot s( t_y, t_y', \fnjp-\fnj, l_y-2^{\fnj-\fnjp}l_y' ) & \\
      \cdot a( t_z, t_z', \fnjp-\fnj, l_z-2^{\fnj-\fnjp}l_z' ), &
      \textrm{ if } \fnj > \fnjp.
    \end{array}
  \right.
\end{eqnarray}
The filters \(a\) and \(s\) are defined by
\begin{eqnarray}
  \label{eq:aasa}
  a( t_1, t_2, j, k ) & = & 
  \int_{\setofreals^3} \mdimdualbasisfn{t_1}{0}{0}(x) 
  \frac{\partial^2}{\partial x^2}
  \mdimbasisfn{t_2}{j}{k}(x) d\tau, \;\;\;\; \textrm{ for } j \geq 0 \\
  \label{eq:aasb}
  a( t_1, t_2, j, k ) & = & 
  \int_{\setofreals^3} \mdimdualbasisfn{t_1}{-j}{k}(x) 
  \frac{\partial^2}{\partial x^2}
  \mdimbasisfn{t_2}{0}{0}(x) d\tau, \;\;\;\; \textrm{ for } j < 0 \\
  \label{eq:aasc}
  s( t_1, t_2, j, k ) & = & 
  \int_{\setofreals^3} \mdimdualbasisfn{t_1}{0}{0}(x) 
  \mdimbasisfn{t_2}{j}{k}(x) d\tau, \;\;\;\; \textrm{ for } j \geq 0 \\
  \label{eq:aasd}
  s( t_1, t_2, j, k ) & = & 
  \int_{\setofreals^3} \mdimdualbasisfn{t_1}{-j}{k}(x) 
  \mdimbasisfn{t_2}{0}{0}(x) d\tau, \;\;\;\; \textrm{ for } j < 0 .
\end{eqnarray}
Filter \(a\) is computed with formulas
\begin{eqnarray}
  a( 0, 0, j, k ) & = & 2^{2j} a_0(k) \textrm{ for } j \geq 0 \\
  a( 0, 0, -1, k ) & = & 4 \sum_{\mu=-m}^m h_\mu 
  a( 0, 0, 0, \mu - k ) \textrm{ for } j = -1 \\
  a( 0, 0, j, k ) & = & 4 \sum_{\mu=-m}^m h_\mu 
  a( 0, 0, j + 1, k - 2^{-j-1} \mu ) \textrm{ for } j < -1 \\
  a( 0, 1, j, k ) & = & a( 0, 0, j+1, 2k+1 ), \textrm{ for } j \geq 0
  \\
  a( 0, 1, -1, k ) & = & 4 a_0(1-k) \\
  a( 0, 1, j, k ) & = & 4 a( 0, 0, j+1, k - 2^{-j-1} ),
  \textrm{ for } j < -1 \\
  a( 1, 0, 0, k ) & = & \sum_{\nu=-m}^m \gtilde_\nu 
  a( 0, 0, -1, \nu - 2k ) \\
  a( 1, 0, j, k ) & = & 4 \sum_{\nu=-m}^m \gtilde_\nu 
  a( 0, 0, j - 1, k - 2^{j-1} \nu ), \textrm{ for } j > 0 \\
  a( 1, 0, j, k ) & = & \sum_{\nu=-m}^m \gtilde_\nu 
  a( 0, 0, j - 1, 2k + \nu ), \textrm{ for } j < 0 \\
  a( 1, 1, j, k ) & = & a( 1, 0, j + 1, 2k + 1 ),
  \textrm{ for } j \geq 0 \\
  a( 1, 1, -1, k ) & = & 4 a( 1, 0, 0, 1 - k ) \\
  a( 1, 1, j, k ) & = & 4 a( 1, 0, j + 1, k - 2^{-j-1} ),
  \textrm{ for } j < -1
\end{eqnarray}
where
\begin{equation}
  a_0(k) := \int_\setofreals
  \dualphi(x) \frac{\partial^2}{\partial x^2}
  \varphi(x - k) dx .
\end{equation}
Filter \(s\) is computed with formulas
\begin{eqnarray}
  s( 0, 0, j, k ) & = & \delta_{k,0}, \textrm{ for } j \geq 0 \\
  s( 0, 0, -1, k ) & = & h_k \\
  s( 0, 0, j, k ) & = & \sum_{\mu=-m}^m h_\mu 
  s( 0, 0, j + 1, k - 2^{-j-1} \mu ), \textrm{ for } j < -1 \\
  s( 0, 1, j, k ) & = & 0, \textrm{ for } j \geq 0 \\
  s( 0, 1, -1, k ) & = & \delta_{k,1} \\
  s( 0, 1, j, k ) & = & s( 0, 0, j + 1, k - 2^{-j-1}), 
  \textrm{ for } j < -1 \\
  s( 1, 0, 0, k ) & = & 0 \\
  s( 1, 0, j, k ) & = & \sum_{\nu=-m}^m \gtilde_\nu
  \delta_{k,2^{j-1}\nu},
  \textrm{ for } j > 0 \\
  s( 1, 0, j, k ) & = & 0, \textrm{ for } j < 0 \\
  s( 1, 1, 0, k ) & = & \delta_{k,0} \\
  s( 1, 1, j, k ) & = & 0, \textrm{ for } j \neq 0
\end{eqnarray}

\section{Constant and Interpolated Pseudopotentials}
\label{sec:potential}

The Coulomb potential arising from a single nucleus is
\begin{equation}
  \label{eq:pot-func}
  V(r) = -\frac{Z}{r} 
\end{equation}
where \(Z\) is the charge of the nucleus.

\begin{figure}
  \centering
  \includegraphics[width=8.6cm]{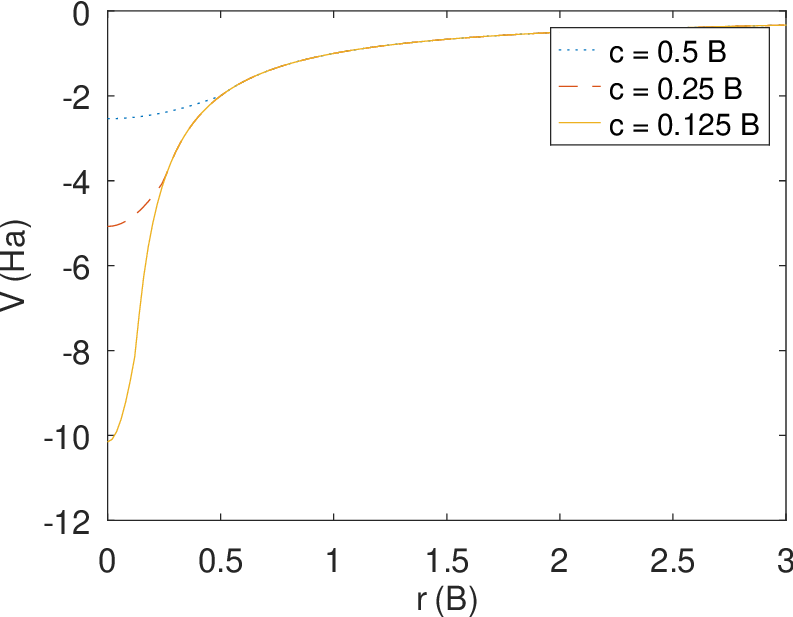}
  \caption{Pseudopotentials $V_{\textrm{interp}}$ with $D = 7$.
    Parameter $c$ is the cutoff value, see section
    \ref{sec:potential}.}
  \label{fig:pseudopotential}
\end{figure}

We avoid the singularity by using a pseudopotential.  We define \(c\)
to be the cutoff point of
the pseudopotential and \(D\) to be the degree of the interpolating
polynomial used in the pseudopotential. Actually we use \(c=2^{-\jmax}\)
where \(\jmax\) is the highest resolution level in the wavelet basis.
Parameter \(D\) has to be an odd integer and we define \(n = (D+1)/2\).
We define
\begin{equation}
  V_1(r) := -\frac{1}{r}, \;\;\;\; r \geq 0,
\end{equation}
\begin{equation}
  \mathbf{s} := (-nc, -(n-1)c, \ldots, -2c, -c, c, 2c, \ldots, (n-1)c,
  nc) ,
\end{equation}
and
\begin{equation}
  \mathbf{t}[i] := V_1(\mathbf{s}[i])
\end{equation}
where \(i=1,\ldots,2n\). Let \(P\) be the interpolating polynomial of
degree at most \(D\) having value \(\mathbf{t}[i]\) at point
\(\mathbf{s}[i]\) for \(i=1,\ldots,2n\).
Now we define the interpolated pseudopotential with
\begin{equation}
  V_{\textrm{interp}}(r) :=
  \left\{
  \begin{array}{ll}
    V_1(r), & r \geq c \\
    P(r),  & r < c
  \end{array}
  \right.
\end{equation}
The actual pseudopotential of a nucleus with charge \(Z\) and location
\(\mathbf{R}\) is then
\begin{equation}
  V(\mathbf{r}) = Z V_{\textrm{interp}}(\abs{\mathbf{r}-\mathbf{R}}) .
\end{equation}
Function \(V_{\textrm{interp}}\) with different values of \(c\) is
plotted in figure \ref{fig:pseudopotential}.
We may also use the cut pseudopotential defined by
\begin{equation}
  V_{\textrm{cut}}(r) :=
  \left\{
  \begin{array}{ll}
    V_1(r), & r \geq c \\
    -\frac{1}{c},  & r < c
  \end{array}
  \right.
\end{equation}
where \(c = 2^{-\jmax-1}\).

\section{HGH Pseudopotential}

By using pseudopotentials the number of computed orbitals can be
reduced and since the pseudo wavefunctions are smoother than
all-electron wavefunctions the basis function set can also be reduced.
In a pseudopotential computation only the valence electrons are
actually computed and the effect of the core electrons is handled by
replacing the nuclear potential with a pseudopotential. We use the
Hartwigsen-Goedecker-Hutter (HGH) pseudopotential
\cite{hgh1998,g2008} in these computations.

The HGH pseudopotential consists of a local and nonlocal part. The
local part is a function of the position as the nuclear potential. The
nonlocal part is a linear operator and it is not a function of
position. The local pseudopotential is given by
\begin{align*}
  V_{\mathrm{loc}}(r) = & \frac{-Z_{\mathrm{ion}}}{r} \erf
  \left( \frac{r}{\sqrt{2} r_{\mathrm{loc}}}\right)
  + \exp \left( - \frac{1}{2} \left(
  \frac{r}{r_{\mathrm{loc}}} \right)^2 \right) \\
  & \times
  \left( C_1 + C_2 \left( \frac{r}{r_{\mathrm{loc}}} \right)^2
  + C_3 \left( \frac{r}{r_{\mathrm{loc}}} \right)^4
  + C_4 \left( \frac{r}{r_{\mathrm{loc}}} \right)^6
  \right)
\end{align*}
where \(r\) is the distance from the nucleus.
Note that
\begin{equation}
  \lim_{r \to 0} \frac{-Z_{\mathrm{ion}}}{r} \erf
  \left( \frac{r}{\sqrt{2} r_{\mathrm{loc}}}\right)
  = \frac{-Z_{\mathrm{ion}}}{r_{\mathrm{loc}}} \sqrt{\frac{2}{\pi}}
\end{equation}
and the local pseudopotential is defined at the origin (nucleus), too.

The nonlocal pseudopotential is defined by
\begin{equation}
  V_{\mathrm{nonlocal}}[\wf] = \vecr \in \setofreals^3 \mapsto \sum_l
  \int V_l(\vecr, \vecrp)
  \wf(\vecrp) d^3 \vecrp
\end{equation}
where
\begin{equation}
  \label{eq:nlbasis}
  V_l(\vecr, \vecrp) = \sum_{i=1}^3 \sum_{j=1}^3 \sum_{m = -l}^l
  Y_{l,m}(\hat{\mathbf{r}}) p^l_i(r) h^l_{i,j} p^l_j(r')
  Y_{l,m}^*(\hat{\mathbf{r}}')
\end{equation}
for each nucleus.
The origin of the coordinate system in \eqref{eq:nlbasis} is located
at the nucleus. The functions \(p^l_i\) are defined by
\begin{equation}
  p^l_i(r) = \frac{\sqrt{2}r^{l+2(i-1)} \exp\left(
    -\frac{r^2}{2r_l^2}\right)}
  {r_l^{l+(4i-1)/2}\sqrt{\Gamma\left(l + \frac{4i-1}{2}\right)}}
\end{equation}
where parameter \(r_l\) is given in Bohrs.
The range of values \(l\) is determined by the actual pseudopotential.
The spherical harmonics \(Y_{l,m}\) in equation \eqref{eq:nlbasis} can
be replaced by orthonormal linear combinations of \(Y_{l,m}\), \(m =
-l, \ldots, l\). This allows us to avoid computation with complex
valued functions.

\section{Atomic and Molecular Orbitals}
\label{sec:orbitals}

\begin{table}
  \caption{\label{tab:bases}Computation grids. We define
  \(\zpm{n} := \{ k \in \setofints : \left\vert k \right\vert \leq n \}\).
  }
  %% \begin{ruledtabular}
  \vspace{0.3cm}
  \begin{tabular}{rl}
    \toprule
    number & grid points \\
    \midrule
    \basisref{1} & \(\frac{1}{2} \zpmthreed{20}\) \\
    \basisref{2} & \(\frac{1}{2} \zpmthreed{20} \cup \frac{1}{4}
    \zpmthreed{10}\) \\
    \basisref{3} & \(\frac{1}{2} \zpmthreed{20} \cup \frac{1}{4}
    \zpmthreed{10} \cup \frac{1}{8} (\zpm{4} \times \zpm{4} \times
    \zpm{10})\) \\
    \basisref{4} & \(\frac{1}{4} \zpmthreed{60}\) \\
    \basisref{5} & \(\frac{1}{2} \zpmthreed{30} \cup \frac{1}{4}
    \zpmthreed{15} \) \\
    \basisref{6} & \(\zpmthreed{38} \cup \frac{1}{2} \zpmthreed{19}\) \\
    \basisref{7} & \(\frac{1}{4} \zpmthreed{40}\) \\
    \basisref{8} & \(\zpmthreed{10} \cup \frac{1}{2} \zpmthreed{10}\) \\
    \basisref{9} & \(\frac{1}{2} \zpmthreed{20} \cup \frac{1}{4}
    \zpmthreed{20}\) \\
    \basisref{10} & \(\frac{1}{4} \zpmthreed{40} \cup \frac{1}{8}
    \zpmthreed{40}\) \\
    \basisref{11} & \(\zpmthreed{10} \cup \frac{1}{2} \zpmthreed{5}\) \\
    \basisref{12} & \(\frac{1}{2} \zpmthreed{20} \cup \frac{1}{4}
    \zpmthreed{10} \cup \frac{1}{8} (\zpm{4} \times \zpm{4} \times
    \zpm{15})\) \\
    \basisref{13} & \(\frac{1}{4} \zpmthreed{40} \cup \frac{1}{8}
    \zpmthreed{20}\) \\
    \basisref{14} & \(\frac{1}{4} \zpmthreed{60} \cup \frac{1}{8}
    \zpmthreed{30}\) \\
    \bottomrule
  \end{tabular}
  %% \end{ruledtabular}
\end{table}

\begin{table}
  \caption{\label{tab:hydrogen-atom}Total energy of the hydrogen atom. Quantity
    \(\spacingsym\) is the distance between grid points in the highest
    resolution level.
    The numbers in column ``basis'' refer to table
    \ref{tab:bases} and ``TH'' means this work.}
  %% \begin{ruledtabular}
  \vspace{0.3cm}
  \begin{tabular}{lrrlr}
    \toprule
    source & basis & \(\spacingsym\; (\bohr)\) & pseudopot.
    & \(E\; (\mathrm{Ha})\) \\
    \midrule
    TH & \basisref{7} & 0.25 & const. & -0.487470 \\
    TH & \basisref{8} & 0.5 & const. & -0.462247  \\
    TH & \basisref{9} & 0.25 & const. & -0.487470 \\
    TH & \basisref{10} & 0.125 & const. & -0.496380 \\
    TH & \basisref{7} & 0.25 & interp. & -0.478328 \\
    TH & \basisref{8} & 0.5 & interp. & -0.439146 \\
    TH & \basisref{9} & 0.25 & interp. & -0.478328 \\
    TH & \basisref{10} & 0.125 & interp. & -0.493471 \\
    TH & \basisref{7} & 0.25 & HGH & -0.499294 \\
    TH & \basisref{8} & 0.5 & HGH & -0.589957 \\
    TH & \basisref{9} & 0.25 & HGH & -0.499295 \\
    TH & \basisref{10} & 0.125 & HGH & -0.499899 \\
    CCCBDB \cite{cccbdb} & & & & -0.466582 \\
    BigDFT \cite{g2008,m2014} & & & HGH & -0.499969 \\
    exact & & & none & -0.5 \\
    \bottomrule
  \end{tabular}
  %% \end{ruledtabular}
\end{table}

\begin{figure}
  \centering
  \includegraphics[width=8.6cm]{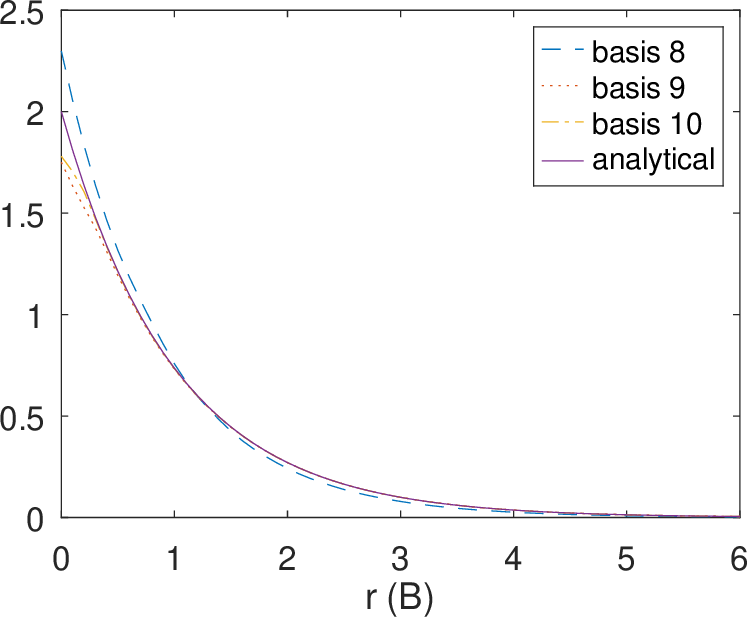}
  \caption{Radially averaged wavefunctions of the hydrogen atom for
    several bases.
    The solid line is the analytical radial wavefunction.
    The basis numbers refer to table \ref{tab:bases}.}
  \label{fig:hydrogen-wavefunctions}
\end{figure}

\begin{figure}
  \centering
  \includegraphics[width=8.6cm]{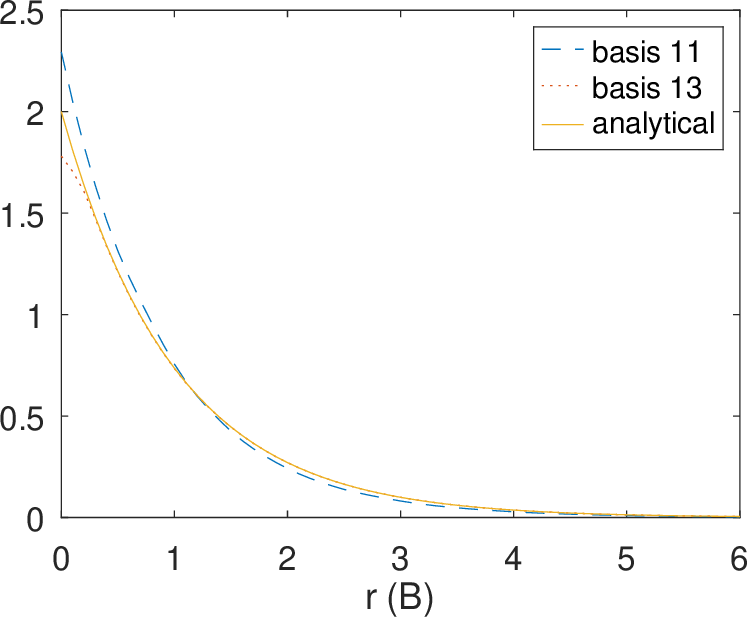}
  \caption{Radially averaged wavefunctions of the hydrogen atom for
    some bases.
    The solid line is the analytical radial wavefunction.
    The basis numbers refer to table \ref{tab:bases}.}
  \label{fig:hydrogen-wavefunctions2}
\end{figure}

\begin{table}
  \caption{\label{tab:hydrogen-exc}Hydrogen atom orbital energies
    with the HGH pseudopotential. We labelled the resulting \(2\mathit{p}\)
    orbitals with a, b, and c. All the orbitals presented are
    approximately orthogonal.}
  %% \begin{ruledtabular}
  \vspace{0.3cm}
  \begin{tabular}{rrr}
    \toprule
    Orbital & Computed energy (Ha) & Exact energy (Ha) \\
    \midrule
    \(1\mathit{s}\) & -0.499295 & -0.5 \\
    \(2\mathit{s}\) & -0.120957 & -0.125 \\
    \(2\mathit{p}_{\mathrm{a}}\) & -0.123045 & -0.125 \\
    \(2\mathit{p}_{\mathrm{b}}\) & -0.123045 & -0.125 \\
    \(2\mathit{p}_{\mathrm{c}}\) & -0.123045 & -0.125 \\
    \bottomrule
  \end{tabular}
  %% \end{ruledtabular}
\end{table}

\begin{table}
  \caption{\label{tab:helium-atom}Energetics of the helium atom. All
    our and BigDFT
    computations use HGH pseudopotential.
    The numbers in column ``basis'' refer to table
    \ref{tab:bases} and ``TH'' means this work.}
  %% \begin{ruledtabular}
  \vspace{0.3cm}
  \begin{tabular}{lrrlrr}
    \toprule
    source & basis & \(\spacingsym \; (\bohr)\) & exchange potential
    & \(E \; (\mathrm{Ha})\) & \(E_{\mathrm{orb}} \; (\mathrm{Ha})\) \\
    \midrule
    TH & \basisref{4} & 0.25 & HF & -2.901959 & -0.971927 \\
    TH & \basisref{5} & 0.25 & HF & -2.901180 & -0.970247 \\
    TH & \basisref{4} & 0.25 & LDA & -2.821511 & -0.629713 \\
    TH & \basisref{5} & 0.25 & LDA & -2.819951 & -0.628152 \\
    TH & \basisref{14} & 0.125 & HF & -2.916129 & -0.974768 \\
    CCCBDB \cite{cccbdb} & & & HF & -2.807584 & \\
    CCCBDB \cite{cccbdb} & & & LSDA & -2.809599 & \\
    BigDFT \cite{g2008,m2014} & & & HF & -2.862303 & \\
    BigDFT \cite{g2008,m2014} & & & LDA & -2.833895 & \\
    HF limit \cite{ff1977} & & & HF & -2.862 & -0.918 \\
    \bottomrule
  \end{tabular}
  %% \end{ruledtabular}
\end{table}

\begin{landscape}
\begin{table}
  \caption{\label{tab:hydrogen-molecule}Hydrogen molecule.
    The numbers in column ``basis'' refer to table
    \ref{tab:bases} and ``TH'' means this work.
  }
  %% \begin{ruledtabular}
  \vspace{0.3cm}
  \begin{tabular}{lrrrllrrr}
    \toprule
    source & basis & \(a \; (\bohr)\) & \(\spacingsym \; (\bohr)\) & pseudopot.
    & exch.
    & \(E_{\mathrm{system}} \; (\mathrm{Ha})\) &
    \(E_{\mathrm{binding}} \; (\mathrm{Ha})\) & \(d \; (\mathrm{B})\) \\
    \midrule
    TH & 1 & 1.0 & 0.5 & interp. & HF & -1.045883 & 0.167601 & 1.855140 \\
    TH & 2 & 1.0 & 0.25 & interp. & HF & -1.156554 & 0.199450 & 1.501870 \\
    TH & 3 & 1.0 & 0.125 & interp. & HF & -1.186176 & 0.210390 & 1.454593 \\
    TH & 4 & 1.0 & 0.25 & HGH & HF & -1.188779 & 0.190189 & 1.397995 \\
    TH & 5 & 1.0 & 0.25 & HGH & HF & -1.187995 & 0.189405 & 1.397991 \\
    TH & 6 & 0.4 & 0.20 & HGH & HF & -1.188547 & 0.189383 & 1.389890 \\
    TH & 4 & 1.0 & 0.25 & HGH & LDA & -1.157528 & 0.158938 & 1.485861 \\
    TH & 5 & 1.0 & 0.25 & HGH & LDA & -1.155960 & 0.157378 & 1.485851 \\
    CCCBDB \cite{cccbdb} & & & & & HF & -1.117506 & 0.184342 & 1.345 \\
    CCCBDB \cite{cccbdb} & & & & & LSDA & -1.157014 & 0.248654 & 1.391 \\
    BigDFT \cite{g2008,m2014} & & & & HGH & HF & -1.133393 & 0.133455 & 1.386175 \\
    BigDFT \cite{g2008,m2014} & & & & HGH & LDA & -1.136870 & 0.136932 & 1.445097 \\
    %% chapter 5.2
    experimental \cite[chapter 5.2]{dommelen} & & & & & & & 0.166 & 1.40 \\
    \bottomrule
  \end{tabular}
  %% \end{ruledtabular}
\end{table}
\end{landscape}

\begin{figure}
  \centering
  \includegraphics[width=8.6cm]{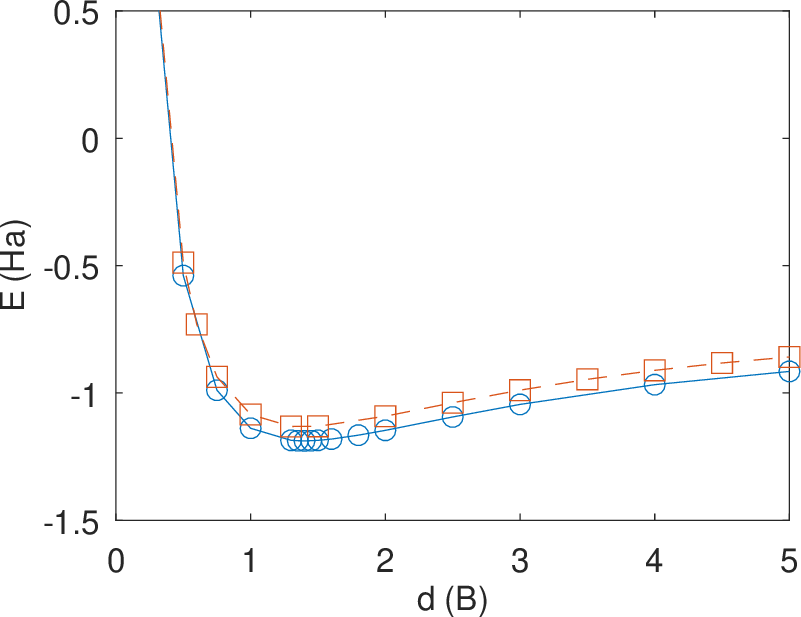}
  \caption{Energy of \(\mathrm{H}_2\) as a function of internuclear
    distance calculated with \hartreefock\ method and HGH
    pseudopotential. The solid line is our computation with
    interpolating wavelets and basis set \basisref{5}, \(g = 0.25 \;
    \bohr\), and the dashed
    line is computed with BigDFT and (fine) grid spacing
    \(g = 0.225 \; \bohr\).}
  \label{fig:h2-dissociation-curve}
\end{figure}

\begin{table}
  \caption{\label{tab:H2-ion}Hydrogen molecule ion $\mathrm{H}_2^+$. The
    analytical results are equal to the experimental results.
    The numbers in column ``basis'' refer to table
    \ref{tab:bases} and ``TH'' means this work.
    Abbreviation ``exp.'' means experimental.
  }
  %% \begin{ruledtabular}
  \vspace{0.3cm}
  \begin{tabular}{lrrlrrr}
    \toprule
    source & basis & \(\spacingsym \; (\bohr)\) & pseudopot.
    & \(E_{\mathrm{system}} \; (\mathrm{Ha})\) &
    \(E_{\mathrm{binding}} \; (\mathrm{Ha})\) & \(d \; (\mathrm{B})\) \\
    \midrule
    TH & \basisref{1} & 0.5 & interp. & -0.520169 & 0.082028 & 2.371005 \\
    TH & \basisref{2} & 0.25 & interp. & -0.573665 & 0.095310 & 2.021654 \\
    TH & \basisref{12} & 0.125 & interp. & -0.589135 & 0.101242 & 2.015143 \\
    TH & \basisref{11} & 0.5 & HGH & -0.712279 & 0.121898 & 2.007951 \\
    TH & \basisref{2} & 0.25 & HGH & -0.601783 & 0.102460 & 2.006845 \\
    TH & \basisref{7} & 0.25 & HGH & -0.601636 & 0.102342 & 2.005329 \\
    TH & \basisref{13} & 0.125 & HGH & -0.602448 & 0.102549 & 1.999338
    \\
    CCCBDB \cite{cccbdb} & & & & -0.582697 & 0.116115 & 2.005 \\
    BigDFT \cite{g2008,m2014} & & & HGH & -0.602489 & 0.102520 & 1.995677 \\
    exp. \cite[chapter 4.6]{dommelen} & & & & & 0.103 & 2.00 \\
    \bottomrule
  \end{tabular}
  %% \end{ruledtabular}
\end{table}

\begin{table}
  \caption{\label{tab:LiH-molecule}Lithium hydride molecule. All
    our and BigDFT computations were done with the HGH
    pseudopotential. The CCCBDB all-electron total energies are not
    presented in the table because the energies calculated by us do
    not include the energies of the core electrons.
    The numbers in column ``basis'' refer to table
    \ref{tab:bases} and ``TH'' means this work.
  }
  %% \begin{ruledtabular}
  \vspace{0.3cm}
  \begin{tabular}{lrrrlrrrr}
    \toprule
    source & basis & \(a \; (\bohr)\) & \(\spacingsym \; (\bohr)\) & exch.
    & \(E_{\mathrm{system}} \; (\mathrm{Ha})\) &
    \(E_{\mathrm{binding}} \; (\mathrm{Ha})\) & \(d \; (\mathrm{B})\) \\
    \midrule
    TH & \basisref{4} & 1.0 & 0.25 & HF & -0.817817 & 0.117472 &
    2.879961 \\
    TH & \basisref{5} & 1.0 & 0.25 & HF & -0.817030 & 0.116686 &
    2.880018 \\
    TH & \basisref{4} & 1.0 & 0.25 & LDA & -0.811482 & 0.111137 &
    3.013411 \\
    TH & \basisref{5} & 1.0 & 0.25 & LDA & -0.809904 & 0.109560 &
    3.013474 \\
    TH & \basisref{6} & 0.4 & 0.2 & HF & -0.817007 & 0.116370 &
    2.863496 \\
    CCCBDB \cite{cccbdb} & & & & HF & & 0.081274 & 2.855 \\
    CCCBDB \cite{cccbdb} & & & & LSDA & & 0.135698 & 2.899 \\
    BigDFT \cite{g2008,m2014} & & & & HF & -0.760938 & 0.059911 & 2.866004 \\
    BigDFT \cite{g2008,m2014} & & & & LDA & -0.776416 & 0.075389 & 2.930745 \\
    \bottomrule
  \end{tabular}
  %% \end{ruledtabular}
\end{table}

\begin{figure}
  \centering
  \includegraphics[width=8.6cm]{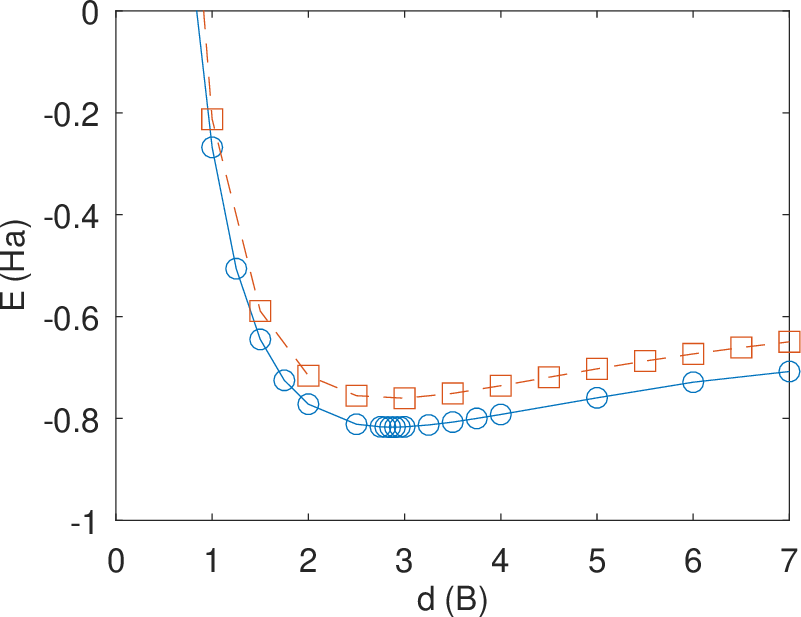}
  \caption{Total energy of the lithium hydride molecule as a function of the
    internuclear distance calculated with the \hartreefock\ method and
    HGH pseudopotential. The solid line has been computed with
    interpolating wavelets and basis set \basisref{5},
    \(g = 0.25 \; \bohr\), and the dashed
    line with BigDFT and (fine) grid spacing \(g = 0.225 \; \bohr\).}
  \label{fig:LiH-energy}
\end{figure}

The hydrogen atom, hydrogen molecule ion, and lithium HGH computations
presented here have been done with Arnoldi method. The helium,
hydrogen molecule, and lithium hydride computations have been done
with self-consistent iteration and Arnoldi method. When there is only
one resolution level in the basis the Hamiltonian and Laplacian
matrices are hermitian and the Arnoldi method reduces to a variant of
the Lanczos method. Furthermore, the
ordinary conjugate gradient method could be used for the Poisson
equation.  Two methods have been used for solving the Poisson equation
for the general case: CGNR and GMRES. We ran a benchmark for these
methods and GMRES was 8 times faster when the accuracies of the
results were approximately the same.  This is because the GMRES code is
parallelized better than CGNR.
All the computations use 8th
order Deslauriers--Dubuc wavelets (polynomial span 7).  We use
interpolating polynomials of degree 7 for the pseudopotentials. In
some computations the Arnoldi method did not find the desired
eigenvalues. This was solved by rising the number of computed
eigenvalues and the number of basis vectors in the Arnoldi method.
The basis function sets (computation point grids) are presented in
table \ref{tab:bases}. The grid spacing for a wavelet basis is
defined to be \(g = 2^{-j_{\mathrm{max}}} a\) where
\(j_{\mathrm{max}}\) is the maximum resolution level in the basis and
\(a\) is the size of one unit in resolution level 0 in the wavelet
basis in atomic units. See section \ref{sec:schrodinger-general}.

The data presented in the tables in this article was obtained from our
own software (denoted by TH), our own computations with BigDFT
\cite{g2008,m2014}, and CCCBDB \cite{cccbdb}. BigDFT is a quantum
mechanical computation package using Daubechies wavelets. CCCBDB is
large database containing atomic and molecular data.
The CCCBDB energies and internuclear distances in this article use the
STO-3G basis set.
Quantity
\(\spacingsym\) is the distance between grid points in the highest
resolution level and quantity \(d\) is the distance between the nuclei
in the result tables.  \(E_{\mathrm{system}}\) is the total energy of
the system and \(E_{\mathrm{binding}}\) is the binding energy.
We used grid spacing \(0.45 \;\bohr\) (finer grid spacing \(0.225
\;\bohr\)) in all our BigDFT computations.
The BigDFT parameters for determining the size of the basis set were
\texttt{crmult=10.0} and \texttt{frmult=16.0}.
Parameter \texttt{crmult} is used to specify the size of the coarse region and
parameter \texttt{frmult} the size of the fine region around atoms.
The computations were
also made with values \texttt{crmult=5.0} and \texttt{frmult=8.0} but
the results did not differ significantly.  Note that the total energy
does not include the energies of the core electrons in the lithium
hydride HGH computations. Computations using our own software use
interpolating wavelets and BigDFT computations orthonormal Daubechies
wavelets.
    
For the molecular computations the energy of the system as a function
of the distance between the nuclei is computed in three points near
the energy minimum and a second degree polynomial is fitted into these
points. The distance between the nuclei is then the minimum point of
the polymial and the energy of the system is computed at the minimum
distance.  We locate the nuclei at points \((0,0,\pm{}\frac{d}{2a})\)
where \(d\) is the distance between the nuclei in Bohrs. When binding
energies of molecules were computed the energies of atoms were usually
computed with the basis set as the molecule. When the basis was
unsymmetric it was modified. For example, grid \(\zpm{4} \times
\zpm{4} \times \zpm{10}\) becomes \(\zpm{4} \times \zpm{4} \times
\zpm{4}\) for the atoms. See the caption of table \ref{tab:bases} for
the definition of \(\zpm{n}\).

The results for hydrogen atom ground state are presented in table
\ref{tab:hydrogen-atom}.  The radially averaged ground state
wavefunctions of the hydrogen atom are plotted in figures
\ref{fig:hydrogen-wavefunctions} and \ref{fig:hydrogen-wavefunctions2}.
A radial average of a
function \(f : \setofreals^3 \to \setofreals\) is computed by
\begin{equation}
  \bar{f}(r) := \frac{1}{4\pi} \int_{\theta=0}^\pi
  \int_{\phi=0}^{2\pi} f(r,\theta,\phi) \sin \theta \; {\rm d} \phi
      {\rm d} \theta
\end{equation}
where \(r \in \left[ 0, \infty \right[\).
As the angular part of an s-type wavefunction is \(\frac{1}{2\sqrt{\pi}}\)
we estimate a radial wavefunction by
\begin{equation}
  \bar{g}(r) := \frac{1}{2\sqrt{\pi}} \int_{\theta=0}^\pi
  \int_{\phi=0}^{2\pi} f(r,\theta,\phi) \sin \theta \; {\rm d} \phi
  {\rm d} \theta .
\end{equation}
The calculation of hydrogen excited
states uses HGH pseudopotential and basis \basisref{9}.
Results are presented in table \ref{tab:hydrogen-exc}.
The names of the excited states were obtained by computing inner
products between the computed states and analytical states.  The
resulting orbitals are approximately orthonormal and the computed 2p
states are approximately linear combinations of the analytical 2p
states. The largest (in absolute value) inner product between
different orbitals is \(\langle \mathrm{2p_a} \vert \mathrm{2p_c} \rangle
= \msci{5.428}{-4}\).  The quality of the linear combinations can be
measured by a quantity \(\sqrt{1-\norm{Pf}_2^2}\) where \(P\) is the
orthogonal projection from \(L^2(\setofreals^3)\) onto the space
spanned by \(\mathrm{2p_x}\), \(\mathrm{2p_y}\), and
\(\mathrm{2p_z}\).
The value of this quantity is \(0.1265\) for all the computed orbitals
\(\mathrm{2p_a}\), \(\mathrm{2p_b}\), and \(\mathrm{2p_c}\).
Analytical expressions for hydrogenic orbitals can be found for
example in Ref. \cite{af2005}.

Helium atom has been computed using the HGH pseudopotential. Results
are presented in table \ref{tab:helium-atom}. Grid spacing
\(\spacingsym = 0.5 \; \bohr\) does not give sensible results with the
HGH pseudopotential.
We suppose that Froese Fischer's results
\cite{ff1977} can be regarded as the \hartreefock\ limit for
helium. The computed total energies of the helium atom are quite good.

We calculated the hydrogen molecule with the interpolated and HGH
pseudopotentials.  The computation results for the hydrogen molecule
are presented in table \ref{tab:hydrogen-molecule}.  The resulting
dissociation curve with the HF method, HGH pseudopotential, and basis
set \basisref{5} is plotted in figure \ref{fig:h2-dissociation-curve}.
When the HGH pseudopotential was used the computation worked for grid
spacing \(g = 0.25 \; \bohr\) but did not work for spacing \(g = 0.5
\; \bohr\).
The minimum energy of the curve is \(E_0 = -1.188 \;\mathrm{Ha}\) and it is
located at internuclear distance \(d_0 = 1.398 \;\bohr\).
The corresponding values for the BigDFT reference curve are
\(E_0 = -1.132 \;\mathrm{Ha}\) and \(d_0 = 1.385 \;\bohr\).
The results for the hydrogen molecule ion are presented in table
\ref{tab:H2-ion}.

The calculations for lithium hydride molecule are presented in
table \ref{tab:LiH-molecule}. Grid spacing \(\spacingsym =
0.5\;\bohr\) did not yield a physical dissociation curve for the HGH
pseudopotential. Neither \(\spacingsym = 0.5\;\bohr\) nor
\(\spacingsym = 0.25\;\bohr\) yielded a physical dissociation curve
for the interpolated pseudopotential.  The dissociation curve of
lithium hydride computed with the \hartreefock\ method and basis set
\basisref{5} is plotted in figure \ref{fig:LiH-energy}.
The minimum energy of the curve is \(E_0 = -0.817 \;\mathrm{Ha}\) and it is
located at internuclear distance \(d_0 = 2.880 \;\bohr\).
The corresponding values for the BigDFT reference curve are
\(E_0 = -0.760 \;\mathrm{Ha}\) and \(d_0 = 2.921 \;\bohr\).

The energies of the hydrogen atom converge to the exact value for the
constant, interpolated, and HGH pseudopotentials with grid spacings
\(g = 0.5\), \(0.25\) and \(0.125\;\bohr\) (table
\ref{tab:hydrogen-atom}). The \hartreefock\ helium computations yield
approximately same results for \(g = 0.25 \;\bohr\) and \(g = 0.125
\;\bohr\) (table \ref{tab:helium-atom}). The
\hartreefock\ computations of the hydrogen molecule using the HGH
pseudopotential give approximately same energies for \(g = 0.25
\;\bohr\) and \(g = 0.2 \;\bohr\) (table
\ref{tab:hydrogen-molecule}). The calculations of the hydrogen
molecule ion using the HGH pseudopotential yield approximately same
energies and internuclear distances for \(g = 0.25 \;\bohr\) and \(g =
0.125 \;\bohr\) (table \ref{tab:H2-ion}). The
\hartreefock\ computations of the lithium hydride molecule give
approximately same results for \(g = 0.25 \;\bohr\) and \(g = 0.2
\;\bohr\) (table \ref{tab:LiH-molecule}).

\section{Conclusions}

We have shown how to solve the wave equations of hydrogen and helium
atoms, hydrogen molecule ion, and hydrogen and lithium hydride
molecules in a three-dimensional interpolating tensor product wavelet
basis.  As far the authors know only Arias \cite{arias1999} and
Engeness and Arias \cite{ea2002} have done this before. However, they
do not use the dual interpolating MRA to evaluate matrix elements. We
do that and it allows us to neglect the overlap integrals of the basis
functions.

It seems to require large basis sets to obtain numerically
good orbitals for quantum physical systems.
%% TARK. seuraava
Roughly, the description requires at least ten thousand basis
functions.  The most accurate computed bond length of the hydrogen
molecule is good and the energy satisfactory.  The most accurate
binding energies and internuclear distances for the hydrogen molecule
ion in table \ref{tab:H2-ion} are very accurate.  The calculations
with the HGH pseudopotential
performed very well with grid spacing \(\spacingsym = 0.25 \; \bohr\)
but not with \(\spacingsym = 0.5 \; \bohr\). The same phenomenon was
observed with BigDFT, too.

We tested the H and He atom computations with BigDFT so that the
coarse grid
spacing was changed from \(0.45\;\bohr\) to \(0.225\;\bohr\). For some
reason we got slightly worse energies.
We also
found that one level basis set can be replaced with considerably
smaller two level basis set without a significant effect on the
results.

Note that having no more than two resolution levels in the basis makes
the computation of the Laplacian operator simpler and faster,
because in that case the \(s(\ldots)\) terms in the Laplacian
operator, equations \eqref{eq:aara}, \eqref{eq:aarb}, and \eqref{eq:aarc},
are either Kronecker deltas or zero.

\printbibliography

\end{document}